# Estimation and Compensation of Process Induced Variations in Nanoscale Tunnel Field Effect Transistors (TFETs) for Improved Reliability

*Abstract*— Tunnel Field Effect Transistors (TFET) have extremely low leakage current, exhibit excellent subthreshold swing and are less susceptible to short channel effects. However, TFETs do face certain special challenges, particularly with respect to the process induced variations in (i) the channel length and (ii) the thickness of the silicon thin-film and the gate oxide. This paper, for the first time, studies the impact of the above process variations on the electrical characteristics of a Double Gate Tunnel Field Effect Transistor (DGTFET). Using two dimensional device simulations, we propose the Strained Double Gate Tunnel Field Effect Transistor (SDGTFET) with high-k gate dielectric as a possible solution for effectively compensating the process induced variations in the on-current, threshold voltage and subthreshold-swing improving the reliability of the DGTFET.

## I. INTRODUCTION

During recent times, many novel devices that utilize band-to-band tunneling for their operation are being actively investigated [1-30]. At reduced device dimensions, the process induced variations assume utmost importance since they lead to a larger relative physical non-uniformity among the devices [30, 31]. In order to employ tunnel devices for CMOS applications at smaller geometries, the fluctuations of the electrical properties of these devices due to process induced variations in (i) the channel length and (ii) the thickness of the silicon thin-film and the gate oxide need to be estimated and kept under control to improve the reliability of these devices. This paper investigates this aspect of the tunnel field effect transistor and also suggests techniques to make the device less susceptible to the process-induced variations.

A number of simulation and experimental results have been reported for tunneling devices using different structures like: TFETs based on nanotubes or nanowires [1]-[7], tunneling devices in bulk silicon [8]-[9], tunneling devices having single, double or multiple gates employing SOI technology [10]-[25], and vertical tunneling devices [26]-[29]. For our study, we have chosen the structure of Double Gate Tunnel Field Effect Transistor (DGTFET) as many of the process steps involved in its fabrication are the same as the Double Gate Field Effect Transistor (DGFET) technology. Continued advancements in DGFET technology may be put to good advantage in improving the performance of a DGTFET.

In this paper, using two-dimensional device simulation, the relative changes in the electrical parameters (on-current, threshold voltage and subthreshold-swing) under a given change in its physical parameters (the channel length and, the thickness of the silicon thin-film and the gate oxide) is computed. The variation in the on-current due to process variations in a DGTFET is found to be quite high and needs to be considerably reduced. Strained Double Gate Tunnel Field Effect Transistor (SDGTFET) with a high-k gate dielectric is suggested as a solution to reduce the on-current variations and obtain an overall improved device performance. The rest of this paper is organized as follows: Section II describes the structure of a DGTFET and the simulation model used in this study. Section III illustrates the methodology that is used in estimating the impact of process induced effects. Section IV presents an estimate of the process induced variations for an SDGTFET. Section V presents the impact of process variations on an SDGTFET with a high-k gate dielectric. Section VI draws important conclusions out of this study.

## II. DEVICE STRUCTURE AND SIMULATION MODEL

The cross-sectional view of the DGTFET used in this study is shown in Fig. 1. The variations of the electrical parameters of the DGTFET are calculated using ATLAS version 1.12.1.R [32] for the nominal parameters given in Table 1 which have been chosen for an optimal device performance [15], [19]. Since the tunneling process is non-local, spatial profile of the energy bands is taken into account including the band gap narrowing effect [32]. It should be noted that we have taken the source and drain doping profiles as abrupt throughout our simulations as in earlier works [15, 18, 19, 27]. A sharp doping profile improves the performance of the device considerably and a reasonably sharp doping profile seems realistic using selective epitaxy [29]. We have used non-local tunneling model in this study and validated using ref. [15].

## III. ESTIMATION OF PROCESS INDUCED VARIATION IN A DGTFET

In order to estimate the impact of the process induced variations on a DGTFET, the dependency of each of the electrical parameters on the physical parameters is computed. The dependency is found by observing the effect on the electrical parameters by varying one of the physical parameters, keeping all other physical parameters of the device fixed to the value shown in Table 1. Using this dependency, we compute the statistical 3-sigma values for the electrical parameters. The 3-sigma variation in an electrical parameter Y with respect to a particular physical parameter $X_i$ having a 3-sigma variation of δX$_i$, is computed as:

$$\% \delta Y|_{\delta X_i} = \frac{1}{Y} X \left( \frac{\partial Y}{\partial X} \right)_i$$ (1)

The contributions to device electrical parameter variations by fluctuations in different physical parameters are taken to be independent of each other. The 3-sigma values of the electrical parameters (Y) with respect to all the physical parameters (X$_i$'s) are computed as [33]:

$$\% \delta Y = \frac{1}{Y} X \sqrt{\sum_i \left( \frac{\partial Y}{\partial X} \right)_i^2 (\partial X_i)^2}$$ (2)

We have considered the following physical parameters to vary due to process induced variations: the thickness of the silicon thin-film, the channel length and the thickness of the gate-oxide. We assume that silicon body thickness varies by 1 nm (10% of nominal silicon body thickness of 10nm), as taken in ref. [34]. Since the gates are automatically aligned with each other and with the source and the drain, a variation of 12% of the nominal channel length (30 nm) is taken as in ref. [35]. Though the variation in thickness of the gate oxide can be controlled within 5% using novel techniques [36-38], we have taken it to vary by ±2 Å as in ref. [35]. In this paper, we have not considered the effect of discrete nature of dopants and their random fluctuations. These effects can impact the variation on the electrical parameters of the DGTFET [39, 40] and can only be done using atomistic simulations.

The electrical parameters that we have considered in this study are: the on-current, the threshold voltage and the subthreshold swing. The on-current is defined as the drain current at a gate voltage of 1.0 V. The threshold voltage is defined as the gate voltage when the drain current reaches $1 \times 10^{-7}$ A/μm. It should be noted that the threshold voltage of a DGTFET can be controlled by changing the work function of the gate. In this work, the threshold voltage has been adjusted to 0.2 V by adjusting the work-function for all the devices so that the threshold voltage variations can be compared around the same nominal value for all the devices. The subthreshold slope of a tunneling device is strongly dependent on the gate voltage. In this paper we use point subthreshold slope to benchmark the performance of a DGTFET [15]. The point sub-threshold slope is defined as:

$$S = \frac{dV_g}{d(\log)_d} \qquad (3)$$

where $V_g$ is the gate voltage and $I_d$ is the drain current.

The electrical characteristics of a DGTFET are not significantly affected by the variations in channel length up to around 25 nm (less than 1% variation in the electrical parameters) since the tunneling phenomenon is confined to a very small region around the source side. This is in contrast to a DGFET where the channel length significantly affects the characteristics of the device [19]. Using two-dimensional device simulation and equation (1), 3-sigma variations in the electrical parameters due to variation in the channel length are computed for a DGFET of similar dimensions (channel length = 30 nm, silicon body thickness = 10 nm, gate oxide thickness = 3 nm, $\Phi_m$= 4.8 eV and threshold voltage = 0.2 V). It is found that, for a 12 % variation in channel length in a DGFET, the on-current varies by 7 %, the threshold voltage varies by 21 % and subthreshold slope varies by 8 %. This shows that a DGFET is adversely affected by variation in channel length due to short channel effects while a DGTFET exhibits a greater tolerance against it.

The results of computation of 3-sigma variations in electrical parameters, for a DGTFET and a DGFET of similar dimension, are shown in Fig. 2. The on-current variation is found to be around 42% for a DGTFET and around 11% for a DGFET. Since the on-current variation in a DGTFET is very high, it needs to be considerably reduced. The 3-sigma variation in the threshold voltage of a DGTFET is found to be around 25%. This is comparable to the 3-sigma variation in threshold voltage in a DGFET of similar dimensions. However, the major contributors to the variations in threshold voltage in a DGTFET are changes in silicon body thickness and gate oxide thickness while in a DGFET the major contributor is the variation in channel length owing to short channel effects.

In a DGTFET, the point subthreshold slope, as defined by equation (3), is strongly dependent on the gate voltage. The point subthreshold slope is quite small (smaller than 60 mV/decade) at low gate voltages and it increases as the gate voltage is increased [15, 19]. However, the point subthreshold slope at a given gate voltage also changes due to the variations in physical parameters, like silicon body thickness or gate oxide thickness. In this paper, we have computed the variation in point subthreshold slope at the gate voltage where the point subthreshold slope is 60 mV/decade for the nominal device. We find that the 3-sigma variation in point subthreshold slope is around 2 % for a DGTFET and around 8 % for a DGFET of similar dimension.

The results shown in Fig. 2 suggest that the on-current variation in a DGTFET is quite high and should be reduced appreciably, and hence in the following sections we explore techniques to reduce the on-current variation.

## IV. PROCESS INDUCED VARIATIONS IN A STRAINED DOUBLE GATE TUNNEL FIELD EFFECT TRANSISTOR (SDGTFET)

Recently, it has been demonstrated that a higher on-current, lower threshold voltage and a better subthreshold swing can be achieved in an SDGTFET compared to a conventional DGTFET [19]. The structure of an SDGTFET is exactly the same as a DGTEFT, except that the silicon body is strained. The strained silicon in an SDGTFET is SiGe free system and can be fabricated using layer transfer techniques [41]-[43]. The strain in the silicon is controlled by the Ge mole fraction in the original SiGe graded buffer over which Si was epitaxially grown.

The presence of strain causes the bandgap and the effective mass of the carriers in silicon to decrease and the electron affinity of silicon to increase. This can be modeled as follows [44]-[46]:

$$\left( \Delta E_g \right)_{sSi} \quad 0.4 \tag{4}$$

$$\left( \Delta E_C \right)_{sSi} \quad 0.57 \tag{5}$$

$$V_T \ln \left( \frac{N_{VsSi}}{N_{VSi}} \right) \approx \frac{m_{hsSi}^*}{m_{hSi}^*}^{\frac{3}{2}} \tag{6}$$

where $x$ is the strain in equivalent Ge mole fraction in the relaxed SiGe buffer layer; $\left( \Delta E_{g,Si} \right)$ is the decrease in the bandgap of silicon due to strain in eV; $\left( \Delta E_C \right)_{sSi}$ is the increase in electron affinity of silicon due to strain in eV; $V_T$ is the thermal voltage; $N_{VSi}$ and $N_{VsSi}$ are the density of states (DOS) in the valence band in the normal and strained silicon, respectively; $m_{hSi}^*$ and $m_{hsSi}^*$ are the hole DOS effective masses in normal and strained-silicon, respectively. Using equations (4) and (5), for a given $x$, the simulator calculates the change in bandgap and the electron affinity for strained silicon. Using the standard values of $N_{VSi}$ and $m_{hSi}^*$ given in [32] and using equation (6), the simulator calculates the change in the effective density of states and the hole DOS effective mass in strained silicon for a given x. The mobility of the carriers also changes in strained silicon [47]. However, in an SDGTFET, mobility plays a limited role when the on-current is limited by tunnel injection. To the best of our knowledge, we do not know of a tunneling model that takes into account the strain in silicon. Therefore, as in ref. 19, we have taken the tunneling parameters in Strained Silicon and all other simulation parameters same as in silicon.

The results of computation of 3-sigma variations of the electrical parameters for an SDGTEFT are plotted in Fig. 3(a)-3(c). Fig. 3(a) shows that the 3-sigma variation in on-current decreases appreciably as we increase the mole fraction of Germanium. The reduction in the variation can be explained using the following relationship between the tunneling current and various device and material parameters [1], [15]:

$$I \propto \exp\left(-\frac{4\sqrt{2}\sqrt{m^*E_g^{*2}}}{3|e|\hbar\left(\Delta\Phi+E_g\right)}\sqrt{\frac{\varepsilon_{Si}}{\varepsilon_{ox}}t_{ox}t_{Si}}\right) \tag{7}$$

where $m^*$ is the effective carrier mass, $E_g$ is the band gap, $\Delta\Phi$ is the energy range over which tunneling can take place, $t_{ox}$, $t_{Si}$, $\varepsilon_{ox}$, and $\varepsilon_{Si}$ are the oxide and silicon film thicknesses and dielectric constants, respectively, $e$ is the electronic charge and $\hbar$ is the Planck's constant [1, 15, 48]. The variation due to silicon body thickness and gate oxide thickness causes variation in the spatial extend of the tunneling region. Using equation (7), the relative change in the tunneling current with change in silicon body thickness and gate oxide thickness can be written as follows (assuming that $\Delta\Phi$, the energy range over which tunneling can take place is independent of small changes in $t_{Si}$ and $t_{ox}$):

$$\frac{1}{I}\left(\frac{\partial I}{\partial t_{Si}}\right) \propto -\left(\frac{2\sqrt{2}\sqrt{m^*E_g^{*2}}}{3|e|\hbar\left(\Delta\Phi+E_g\right)}\sqrt{\frac{\varepsilon_{Si}}{\varepsilon_{ox}}\frac{t_{ox}}{t_{Si}}}\right) \tag{8}$$

$$\frac{1}{I}\left(\frac{\partial I}{\partial t_{ox}}\right) \propto -\left(\frac{2\sqrt{2}\sqrt{m^*E_g^{*2}}}{3|e|\hbar\left(\Delta\Phi+E_g\right)}\sqrt{\frac{\varepsilon_{Si}}{\varepsilon_{ox}}\frac{t_{Si}}{t_{ox}}}\right) \tag{9}$$

These relationships show that with the reduction in the band-gap of the material, the 3-sigma variation in on-current is expected to decrease both due to the silicon body thickness and gate oxide thickness. Hence, an SDGTFET shows an overall reduction of variation in the on-current. Another interesting observation can be made by examining equations (8) and (9). The variation in on-current is dependent on gate oxide thickness and silicon body thickness. However, any attempt to reduce the variation in on-current by changing these device parameters may improve only one of the components i.e. variation due to silicon body thickness (equation (8)) or variation due to gate oxide thickness (equation (9)). The other component will be adversely affected by this change. The important electrical parameters of the device such as on-current, threshold voltage, subthreshold swing and leakage current are strongly dependent on these physical parameters. Therefore, we have not tried to find an optimum value for these parameters which could minimize the impact of process induced variations. Rather, we have chosen those values of the parameters that are optimal for on-current, threshold voltage, subthreshold swing and leakage current. It should also be pointed out that, throughout this paper, the variation due to silicon body thickness is far greater than due to the gate oxide thickness. The computed 3-sigma variations in the electrical parameters are dependent on the choice of the nominal device parameters. In this paper, we have taken 10 nm as the silicon body thickness, which is rather thin and 3 nm as the gate oxide thickness, which may be considered a thick gate oxide. There are several other simulations as well as experimental studies which have chosen the device parameters similar to the one used in this paper [15, 16, 19, 23, 24, 48]. This choice of parameters of silicon body thickness and gate oxide thickness may be leading to a higher impact of silicon film thickness. However, we believe that the nominal device parameters chosen in this study could be relevant in the fabricated device. A thin silicon body thickness is required to have a good on-current and a maximum current is reached when silicon film thickness is around 10 nm [15].

Figure 3(b) shows that the variation in threshold voltage decrease with increasing Ge mole fraction (from 25% at Ge mole fraction = 0.0 to 17% at Ge mole fraction = 0.5). This trend can be explained using equations (8) and (9). By the definition of threshold voltage, the drain current at threshold voltage for the nominal device is $1x10^{-7}$ A/μm. Due to process induced effects, when the silicon body thickness or the gate oxide thickness changes, the drain current deviates from its nominal value of $1x10^{-7}$ A/μm. The change in threshold voltage, due to change in silicon body thickness or gate oxide thickness, is the change in gate voltage that is required to compensate the above mentioned change in the drain current and bring it back to $1x10^{-7}$ A/μm. However, as equations (8) and (9) show, deviation in drain current would be smaller for a device with smaller bandgap, and hence lesser compensation is required for a device with higher Ge mole fraction. Additionally, since the increase in drain current with the gate voltage is steeper for an SDGTFET with higher Ge mole fraction [19], for the same amount of change in drain current smaller change in the gate voltage is required at a higher Ge mole fraction. Therefore, a small change in the gate voltage can compensate a larger change in drain current in a device with higher Ge mole fraction. Hence, we observe a smaller variation in threshold voltage at a higher Ge mole fraction.

Figure 3(c) shows the variations in point subthreshold slope at different Ge mole fractions. We find that the 3-sigma variations in the point subthreshold slope are quite low (around 2%) and it remains within 2% with increasing Ge mole fraction.

## V. Process induced variations in a SDGTFET with high k dielectric

Equation (8) and (9) show that both the components of 3-sigma variation in on-current i.e. variations due to silicon body thickness and gate oxide thickness, can be reduced simultaneously by increasing the dielectric constant of the gate. Therefore, we examine the effect of replacing gate dielectric in an SDGTFET from the conventional $SiO_2$ ($\varepsilon$=3.9) to $Si_3N_4$ ($\varepsilon$=7.5) and $HfO_2$ ($\varepsilon$=21). We have kept the physical thickness of the gate oxides same for all the devices. The 3-sigma variation in the on-current and the threshold voltage are shown in Fig. 4(a) and Fig. 4(b), respectively. We find that the variation in on-current comes down below 20% when $HfO_2$ is used as a gate dielectric in an SDGTFET. As expected, the variations in threshold voltage also improve for a high-k gate dielectric. Hence it can be concluded that an SDGTFET with a high-k dielectric has a significantly improved tolerance against the process induced variations.

## VI. CONCLUSION

This paper has illustrated the estimation of process-induced variations in a DGTFET. It is shown that the impact of variations in (i) the channel length and (ii) the thickness of the silicon thin-film and the gate oxide on the electrical properties of a DGTFET is quite high and can be one of the limiting factors for its wide-scale application. It is demonstrated that an SDGTFET with a high-k gate dielectric can successfully bring down the impact of process induced variations on the on-current, threshold voltage and subthreshold-swing improving the reliability of the DGTFET in future CMOS applications [43].

**Table 1: Device Parameters used in simulation of an SDGFET**

| Device Parameter | Value |
|---|---|
| Source Doping (atoms/cm$^3$) | $1 \times 10^{20}$, p-type |
| Drain Doping (atoms/cm$^3$) | $5 \times 10^{18}$, n-type |
| Channel Doping (atoms/cm$^3$) | $1 \times 10^{17}$, p-type |
| Channel Length, L (nm) | 30 |
| Gate Oxide Thickness, $t_{ox}$ (nm) | 3.0 |
| Silicon Body Thickness, $t_{Si}$ (nm) | 10 |
| Drain Bias, $V_{DS}$ (V) | 1.0 |
| Thin Film Body Material | Si ($\varepsilon$=11.9) |
| Gate Dielectric Material | SiO$_2$($\varepsilon$=3.9) |

# <u>List of Figures</u>

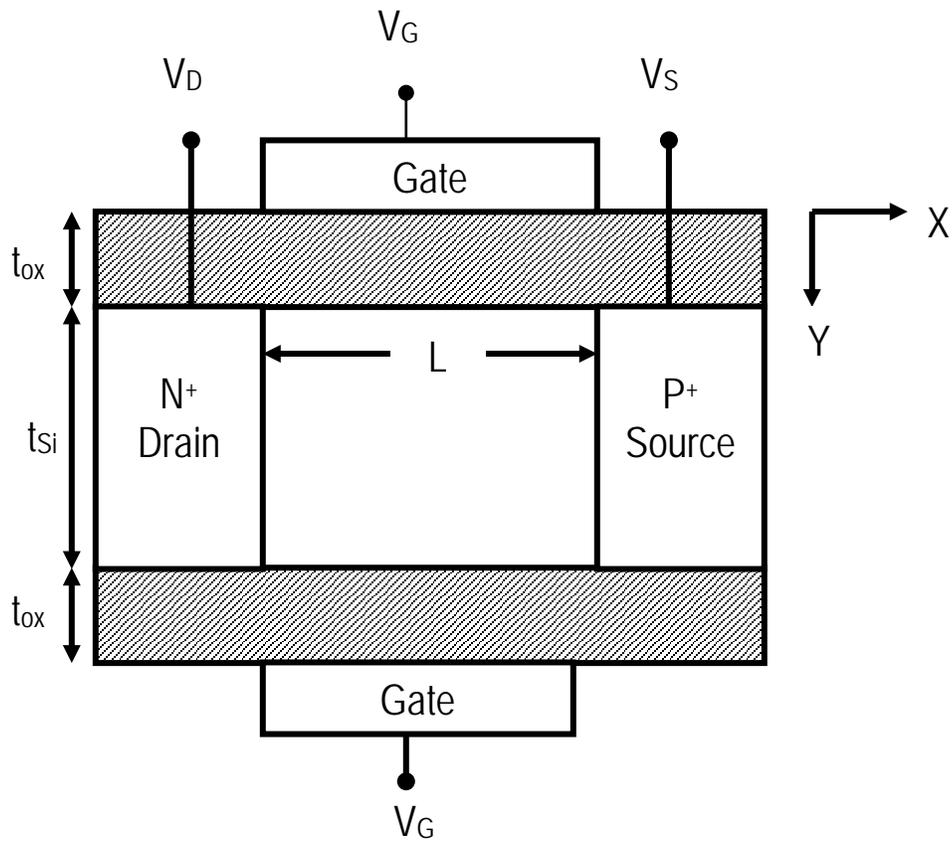

**Fig. 1**

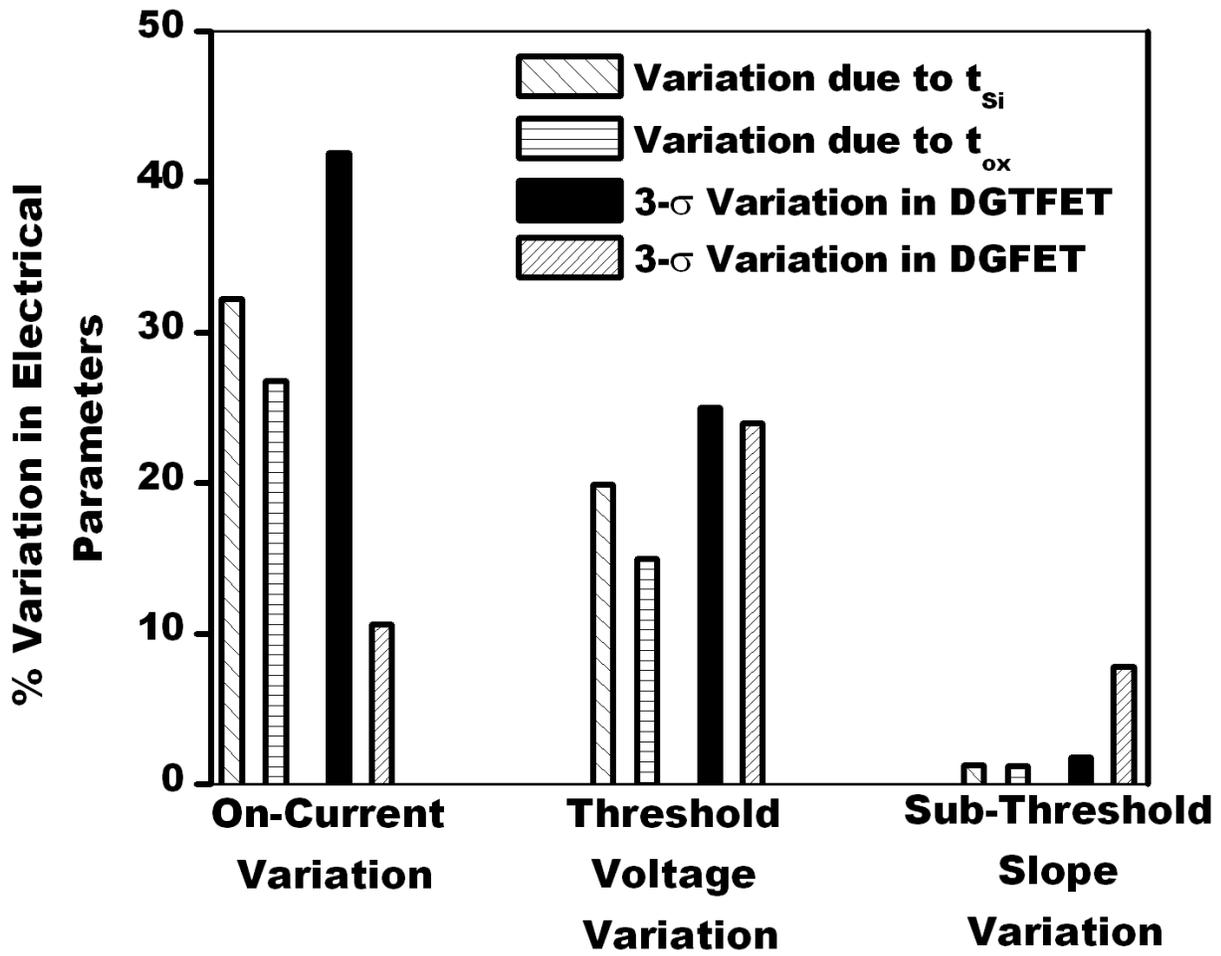

**Fig. 2**

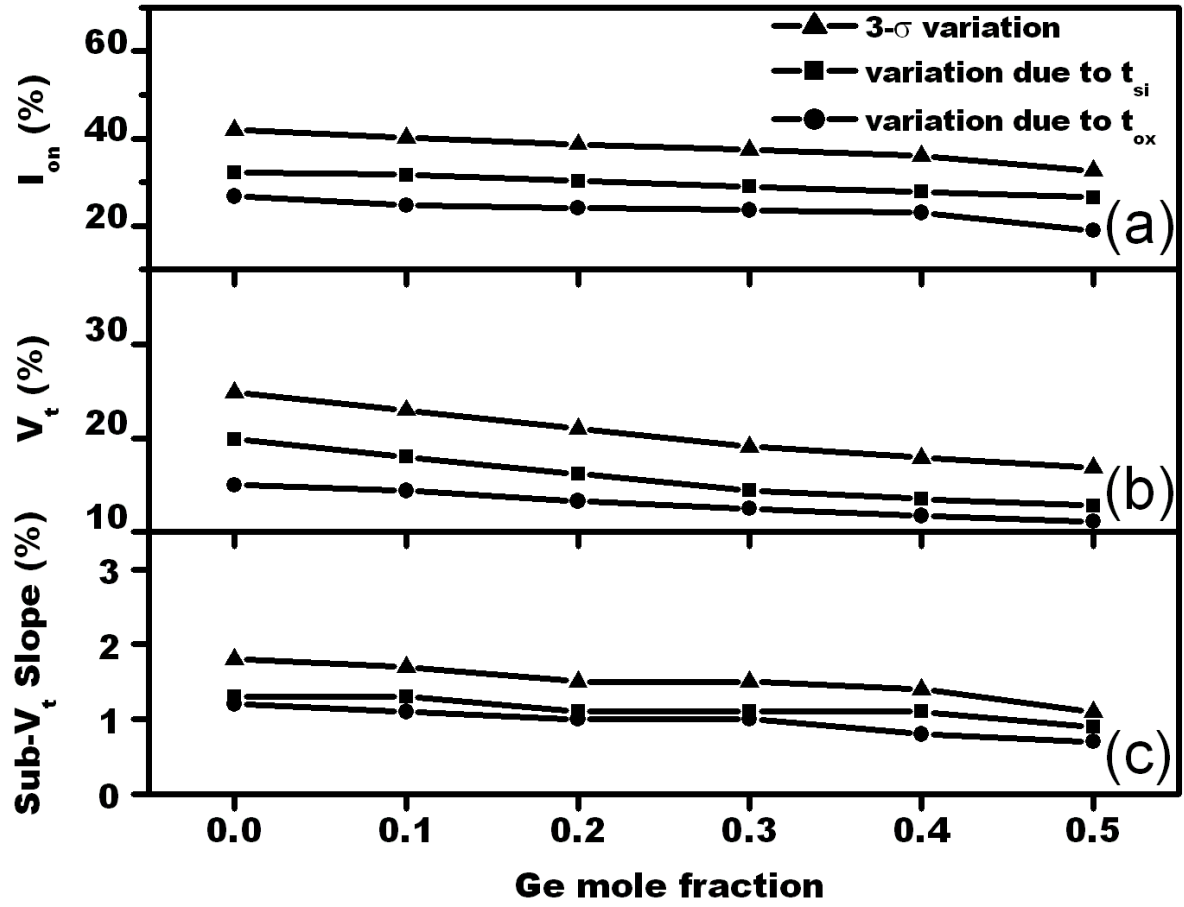

**Fig. 3**

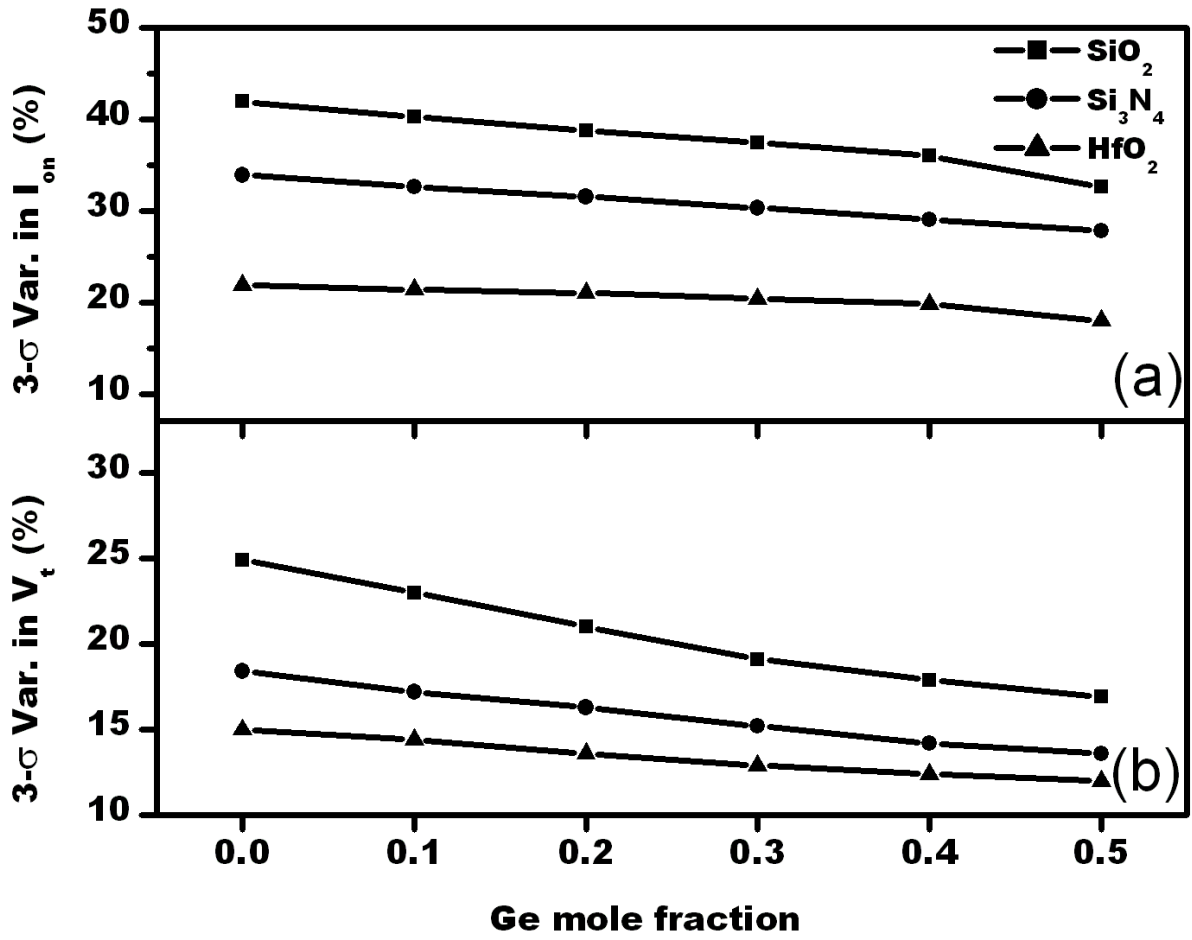

**Fig. 4**

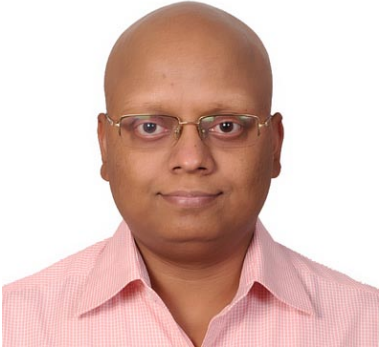

**Sneh Saurabh** received the B. Tech. degree in Electrical Engineering from Indian Institute of Technology, Kharagpur in 2000. Since then he has been working on various aspects of design and verification of integrated circuits. His research interests include quantum devices, low-power methodologies and design optimizations. He is currently pursuing his PhD at Indian Institute of Technology, New Delhi in the field of tunnel devices for CMOS applications.

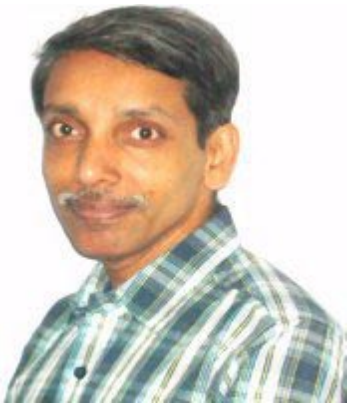

M. Jagadesh Kumar (was born in Mamidala, Andhra Pradesh, India. He received the M.S. and Ph.D. degrees in electrical engineering from the Indian Institute of Technology (IIT), Madras, India.

From 1991 to 1994, he performed a postdoctoral research on the modeling and processing of highspeed bipolar transistors with the Department of Electrical and Computer Engineering, University of Waterloo, Waterloo, ON, Canada. While with the University of Waterloo, he also did research on amorphous-silicon thin-film transistors. From July 1994 to December 1995, he was initially with the Department of Electronics and Electrical Communication Engineering, IIT, Kharagpur, India, and then, he was with the Department of Electrical Engineering, IIT, New Delhi, India, where he became an Associate Professor in July 1997 and has been a Full Professor in January 2005. He is currently the Chair Professor of the NXP (Philips) (currently, NXP Semiconductors India Pvt. Ltd.) established at IIT Delhi by Philips Semiconductors, The Netherlands. He is the Coordinator of the Very Large Scale Integration (VLSI) Design, Tools, and Technology interdisciplinary program at IIT Delhi.

His research interests include nanoelectronic devices, device modeling and simulation for nanoscale applications, integrated-circuit technology, and power semiconductor devices. He has published extensively in these areas of research with three book chapters and more than 145 publications in refereed journals and conferences. His teaching has often been rated as outstanding by the Faculty Appraisal Committee, IIT Delhi.

Dr. Kumar is a fellow of the Indian National Academy of Engineering, The National Academy of Sciences and the Institution of Electronics and Telecommunication Engineers (IETE), India. He is recognized as a Distinguished Lecturer of the IEEE Electron Devices Society (EDS). He is a member of the EDS Publications Committee and the EDS Educational Activities Committee. He is an Editor of the IEEE TRANSACTIONS ON ELECTRON DEVICES. He was the lead Guest Editor for the following: 1) the joint special issue of the IEEE TRANSACTIONS ON ELECTRON DEVICES and the IEEE TRANSACTIONS ON NANOTECHNOLOGY (November 2008 issue) on Nanowire Transistors: Modeling, Device Design, and Technology and 2) the special issue of the IEEE TRANSACTIONS ON ELECTRON DEVICES on Light Emitting Diodes (January 2010 issue). He is the Editor-in-Chief of the IETE Technical Review and an Associate Editor of the Journal of Computational Electronics. He is also on the editorial board of Recent Patents on Nanotechnology, Recent Patents on Electrical Engineering, Journal of Low Power Electronics, and Journal of Nanoscience and Nanotechnology. He has reviewed extensively for different international journals.

He was a recipient of the 29th IETE Ram LalWadhwa GoldMedal for his distinguished contribution in the field of semiconductor device design and modeling. He was also the first recipient of the India Semiconductor Association–VLSI Society of India TechnoMentor Award given by the India Semiconductor Association to recognize a distinguished Indian academician for playing a significant role as a Mentor and Researcher. He is also a recipient of the 2008 IBM Faculty Award. He was the Chairman of the Fellowship Committee of The Sixteenth International Conference on VLSI Design (January 4–8, 2003, New Delhi, India), the Chairman of the Technical Committee for High Frequency Devices of the International Workshop on the Physics of Semiconductor Devices (December 13–17, 2005, New Delhi), the Student Track Chairman of the 22nd International Conference on VLSI Design (January 5–9, 2009, New Delhi), and the Program Committee Chairman of the Second International Workshop on Electron Devices and Semiconductor Technology (June 1–2, 2009, Mumbai, India).